\documentclass[graybox]{svmult}

\usepackage{mathptmx}       
\usepackage{helvet}         
\usepackage{courier}        
\usepackage{type1cm}        
\usepackage{makeidx}         
\usepackage{graphicx}        
\usepackage{multicol}        
\usepackage[bottom]{footmisc}

\makeindex             

\usepackage{amsmath,amssymb}

\DeclareMathOperator{\Var}{Var}

\newcommand{\Varb}[1]{\ensuremath{\Var\!\left[#1\right]}}

\newcommand{\norm}[1]{\left\lVert #1 \right\rVert}

\usepackage{url}

\usepackage{xparse}
\usepackage{ifthen}
\DeclareDocumentCommand{\comment}{o m o o o o}
{\ifthenelse{\draft=1}{
  \IfValueT{#1}{
      \textcolor{red}{\textbf{C (#1) : }#2}
      \IfValueT{#3}{\textcolor{blue}{\textbf{A1 : }#3}}
      \IfValueT{#4}{\textcolor{ForestGreen}{\textbf{A2 : }#4}}
      \IfValueT{#5}{\textcolor{red!50!blue}{\textbf{A3 : }#5}}
      \IfValueT{#6}{\textcolor{Aquamarine}{\textbf{A4 : }#6}}
    }
    \IfNoValueT{#1}{
      \textcolor{red}{\textbf{C : }#2}
      \IfValueT{#3}{\textcolor{blue}{\textbf{A1 : }#3}}
      \IfValueT{#4}{\textcolor{ForestGreen}{\textbf{A2 : }#4}}
      \IfValueT{#5}{\textcolor{red!50!blue}{\textbf{A3 : }#5}}
      \IfValueT{#6}{\textcolor{Aquamarine}{\textbf{A4 : }#6}}
    }
 }{}
}

\usepackage[margin=2cm]{geometry}

\begin{document}

\title*{Unveiling Co-evolutionary Patterns in Systems of Cities: A Systematic Exploration of the SimpopNet Model}
\titlerunning{Unveiling co-evolutionary patterns} 
\author{Juste Raimbault}
\institute{Juste Raimbault \at CASA, UCL and UPS CNRS 3611 ISC-PIF, \email{juste.raimbault@polytechnique.edu}
}
%
%
\maketitle

\abstract{Co-evolutionary processes are according to the evolutionary urban theory at the center of urban systems dynamics. Their empirical observation or within models of simulation remains however relatively rare. This chapter is focused on the co-evolution of transportation networks and cities and applies high performance computing numerical experiments to the SimpopNet co-evolution model in order to understand its behavior. We introduce specific indicators to quantify trajectories of such models for systems of cities, and apply these to exhibit co-evolutionary regimes of the model. This illustrates how the systematic exploration of a simulation model can qualitatively transform the knowledge it provides.\medskip
}

\section{Introduction}

\subsection{Exploring models of simulation}

The development of new knowledge production practices, in particular the use of simulation models to understand complex systems, has been more and more fostered by the increase in computational possibilities. \cite{arthur2015complexity} has proposed in that sense that these trends would consist in a computational shift of science, moving progressively from analytical-based approaches to simulation-based approaches. The study of complexity is naturally not the only field of science benefiting from these technological advances, as witness for example recent progresses in computer vision thanks to deep learning techniques made efficient by intensive computing \cite{lecun2015deep}, or the importance of cloud computing for processing the massive amount of data produced by the LHC detectors \cite{bird2011computing}. These new methods, tools and practices are however particularly suited to the study of complex systems, because among other reasons their flexibility to take into account numerous interacting heterogeneous agents composing this kind of systems. In the case of socio-technical systems, several examples of such streams of research can be given such as generative social science \cite{epstein2006generative}, geosimulation \cite{benenson2004geosimulation}, sociophysics \cite{galam2008sociophysics} or econophysics \cite{mantegna1999introduction}. In that case, models are a crucial piece among other knowledge domains \cite{raimbault2017applied} such as theoretical and empirical investigations. All knowledge domains are how- ever complementary and often necessary, and \cite{raimbault2016cautious} recalls the risks of falling into blind fully computational practices.

The study of urban systems, which are a typical illustration of such complex systems \cite{batty2007cities}, has witnessed a significant gain of knowledge from the ``\textit{liberation of modeling and simulation practices}'' as \cite{banos2013pour} puts it. Recent developments around the evolutive urban theory \cite{pumain1997pour}, synthesized in particular in \cite{pumain2017urban}. Following \cite{banos2017knowledge}, these efforts are the archetype of the complementarity of knowledge domains mentioned above, and have acted as a ``knowledge accelerator'' with a true beneficial interdisciplinary exchange between computer science and geography.

In particular, the development of new methods for model exploration such as Pattern Space Exploration algorithm \cite{cherel2015beyond} or the Calibration Profile algorithm \cite{reuillon2015new}, particularly designed for the use of high performance computing, have allowed a qualitative shift in the knowledge that could be extracted from a simulation model. Several illustrations can be given. A set of parameters that are necessary and sufficient to obtain targeted stylized facts for the emergence of a system of cities are obtained for the SimpopLocal model \cite{pumain2017evaluation}. \cite{arduin2018modelisation} obtains confidence intervals for the estimation of parameters of a non-tractable epidemiological model through the use of the Calibration Profile algorithm. \cite{brasebin2017apports} facilitates urban plannig by exploring the feasible space of building envelopes under the constraints of local regulations. \cite{raimbault2018indirect} indirectly quantifies interactions between networks and territories through the calibration of a simulation model with a genetic algorithm. These results are obtained thanks to the use of the model exploration software OpenMOLE~\cite{reuillon2013openmole}, which is built around three complementary axis: (i) the possibility to embed almost any model as a black box whatever the language in which it is written (as soon as it runs on a linux machine); (ii) the implementation of innovative model exploration and calibration methods; and (iii) a transparent access to high performance computing environments. These features are integrated seamlessly with the use of a specific domain specific language to compose experiment workflows \cite{passerat2017reproducible}.

We consider thus that a considerable gain in knowledge can be observed, from the conceptual or thematic description of a model, to its mathematical formalization, its implementation, its systematic exploration, up to its exploration in deep with the help of specific meta-heuristics. These changes may furthermore be of a qualitative nature, in the sense that the nature of knowledge follows abrupt transitions during the advance of the investigation in this continuum.

The objective of this chapter is to illustrate the impact of these new methods in the case of a model of co-evolution between cities and transportation networks, the SimpopNet model, introduced by \cite{schmitt2014modelisation}. Our contribution is significant on the following points: (i) we provide a supplementary proof-of-concept on the role of new simulation practices, tools and methods; (ii) we introduce a set of indicators to study the behavior of simulation models for systems of cities; (iii) we establish the behavior of this particular model, in particular we assess its sensitivity to spatial initial configuration and unveil the different regimes for interactions between cities and networks it can produce.

The rest of this chapter is organized as follows: we first briefly review co- evolutionary models for systems of cities and describe the model studied. We then introduce methodological elements to study such kind of models for systems of cities, describe results of the systematic exploration, and finally discuss the implication of these.

\subsection{Co-evolution within systems of cities}

Co-evolution within system of cities, in the sense of complex intricate dynamics in space and time, is a central feature of the evolutive urban theory \cite{pumain2010theorie}. \cite{paulus2004coevolution} has for example applied this concept to the study of economic trajectories of French urban areas. Evolutionary economic geography has also developed an extensive literature using this concept for spatial economic systems \cite{schamp201020}, for example for the location of firms and networks \cite{doi:10.1080/00343400802662658}. We use the definition of this concept proposed by \cite{raimbault:tel-01857741}, which can be synthesized as the statistical existence of causal relationships within spatio-temporal niches, for which a practical characterization method uses a weak causality based on lagged correlations \cite{raimbault2017identification}.

Considering more precisely the co-evolution of transportation networks and territories, which is of particular interest because of potential ``structuring effects'' of transportation infrastructures~\cite{pumain2014effets}, some empirical investigations have been proposed by \cite{bretagnolle2003vitesse} and \cite{bretagnolle:tel-00459720} for the French system of cities. The validity of these results was however recently questioned by more thorough data analysis in \cite{mimeur:hal-01616746} and \cite{2018arXiv180409430R}. For this reason, models of co-evolution are crucial to gain further insight into this concept.

These kind of models are however rare for cities and transportation networks, as \cite{raimbault2017models} suggested that this could be due to the fact that this object of study is at the crossroad of several disciplines with different interests and underlying questions. We can give a few examples of such models (see \cite{raimbault:tel-01857741} for a more thorough review). At microscopic and mesoscopic scales, \cite{achibet2014model} describes a model of the co-evolution of buildings and roads, whereas \cite{2018arXiv180505195R} develops a morphogenesis model coupling multi-modeling of road network growth with a reaction-diffusion model for population density. At the scale of systems of cities, \cite{baptistemodeling} proposed a reinforcement co-evolution model. \cite{blumenfeld2010network} has focused on topological breakdown of the network of cities. The SimpopNet model introduced by~\cite{schmitt2014modelisation}, is to the best of our knowledge the only co-evolution model in the perspective of the evolutive urban theory. 

This last model was however not systematically explored, and the question remains if it actually produces patterns of co-evolution at an aggregated level. This makes it a good candidate for our approach. We will in the next section briefly recall the structure of this model.

\subsection{Description of the SimpopNet model}

We reformulate here the SimpopNet model~\cite{schmitt2014modelisation}, following the notations for the formalization of the interaction model introduced by \cite{raimbault2018indirect}, since a certain number of parameters and processes are similar. Cities grow following a specification for their populations $\mu_i(t)$ such that

\begin{equation}
\mu_i(t+1) - \mu_i (t) = \mu_i (t) \cdot \frac{d_G^{\gamma_G}}{N} \sum_{j} \frac{V_{ij}}{<V_{ij}>}
\end{equation}

where the potential $V_{ij}$ is of the form 

\begin{equation}
V_{ij} = \mu_j / d_{ij}^{\gamma_G}
\end{equation}

such that $V_{ii}=0$, $\gamma_G$ is a parameter for the distance decay (which gives indeed a level of hierarchy as a function of distance) and $d_G$ a shape parameter for the decay function which gives the typical distance of interaction.

The network growths at each time step through a process that can be seen as a potential breakdown (as described by \cite{raimbault:tel-01857741}):
\begin{enumerate}
	\item two cities are chosen, the first according to populations with a hierarchy $\gamma_N$ (i.e. with a probability proportional to ${\mu_i}^{\gamma_N}$) and the second following interaction forces $\mu_i \mu_j / d_{ij}^\beta$ with the same hierarchy $\gamma_N$;
	\item a link is then created if the network is not efficient enough given a threshold parameter $\theta_N$, i.e. if $d_{ij}/d^{(N)}_{ij}> \theta_N$;
	\item the links created at a date $t$ have a speed $v(t)$, which will depend on current transportation technologies;
	\item a creation of new intersections to yield a planar graph is done, but only for links with a similar speed.
\end{enumerate}

In order to study a stylized version of the model, we consider a configuration such that $v(t > 0) = v_0$ and $v(0) = 1$ (the initial model considers three values for speed that correspond to the reality of transportation technologies between 1830 and 2000). Indeed, the initial precision in the parametrization of dates and speeds makes it a hybrid model, and should correspond to an application on a real spatial configuration. In a synthetic configuration as used in the model, these parameters have a sense only if we know the behavior of simulated dynamics, and in particular the role of the spatial configuration, i.e. if we are able to differentiate effects linked to the dynamics from effects linked to the initial spatial configuration.

\section{Methods}

\subsection{Generation of synthetic setup configurations}

An important aspect for studying the behavior of such a simulation model is the role of the initial spatial configuration in emerging patterns observed. We therefore apply the methodology developed by \cite{cottineau2017initial}, which allows to extend the sensitivity analysis of a model to spatial meta-parameters (in our case a meta-parameter is a parameter allowing to generate an initial configuration upstream of the model).

A synthetic system of cities is constructed the following way (see \cite{raimbault2016generation} for the notion of synthetic data, calibrated at the first and the second order). A fixed number $N$ of cities is uniformly distributed in space, under the constraint of a minimal distance between each, and their population is attributed following a rank-size law which parameters $P_{m}$ and $\alpha$ can be adjusted (the distribution of city sizes in the original model corresponds to $\alpha\simeq 0.68$ with $R^2=0.98$).

A skeleton of network is created by progressive connection (similar to a percolation algorithm): the algorithm connects cities two by two by closest neighbour in terms of euclidian distance, and then iteratively selects randomly a cluster and connects it perpendicularly to the closest link outside the cluster. The network is then extended by the creation of local shortcuts, through a repetition $n_s$ times of the random selection of a city with a probability proportional to populations, and its connection to a random neighbour in a radius $r_s$ under conditions of a maximal degree $d_s$. The final network is then made planar.

\begin{figure}
	\centering
	\includegraphics[width=0.49\textwidth]{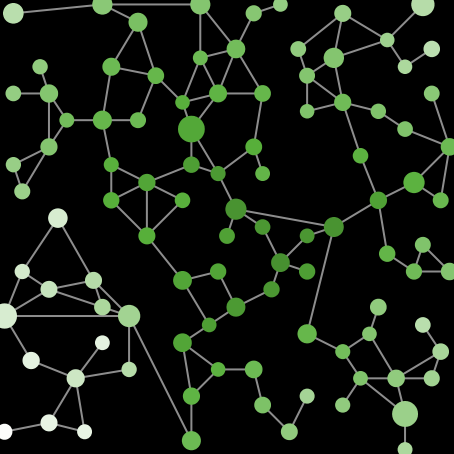}\hspace{0.1cm}
	\includegraphics[width=0.49\textwidth]{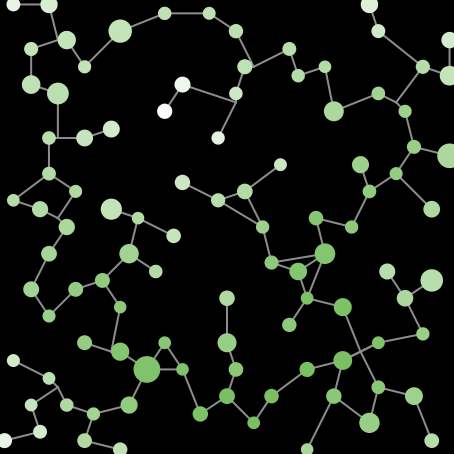}\\\vspace{0.1cm}
	\includegraphics[width=0.49\textwidth]{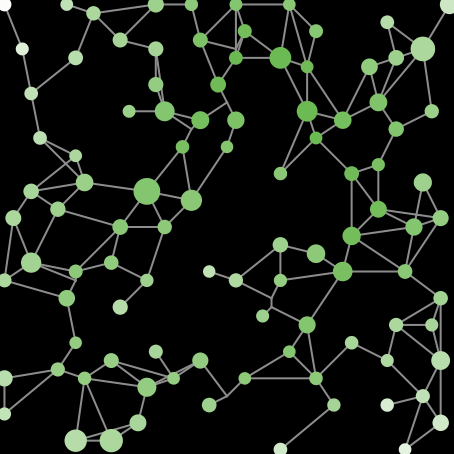}\hspace{0.1cm}
	\includegraphics[width=0.49\textwidth]{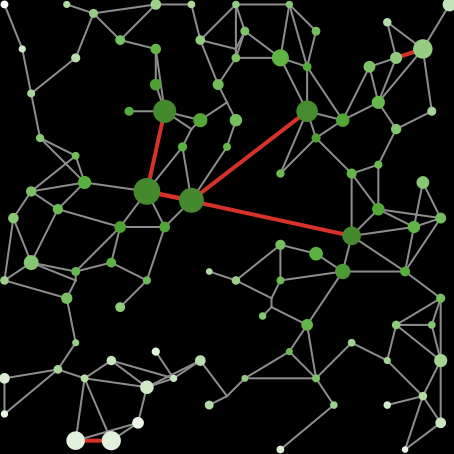}
	\caption{\textbf{Examples of model setup.} (\textit{Top Left}) Unique stylized setup provided in the original implementation of the model \cite{schmitt2014modelisation}; (\textit{Top Right}) Example of a synthetic setup (generator parameters $N=100$, $\alpha=0.8$, $P_m =1.57e5$, $n_s = 10$, $r_s=7$, $d_s = 5$) with a network close to a tree, obtained with a very low number of shortcuts; (\textit{Bottom Left}) Example of a synthetic setup (generator parameters $N=100$, $\alpha=0.8$, $P_m =1.57e5$, $n_s = 70$, $r_s=8$, $d_s = 5$) with a higher number of local connections; (\textit{Bottom Right}) Outcome of the model after $t_f = 100$ time steps on the second synthetic configuration. City size gives the population and color the closeness centrality.\label{fig:exsetup}}
\end{figure}

This process creates networks that visually correspond (in terms of the order of magnitude of the number of loops, and their spatial range) to the initialization of the model, knowing that a single instance of the network does not allow to determine distributions of topological parameters for which a more precise calibration could be done. We show in Fig.~\ref{fig:exsetup} examples of synthetic setups compared to the original model setup.

\subsection{Indicators for trajectories of systems of cities}

\subsubsection{Context}

A crucial aspect of the study of simulation models is the definition of relevant indicators, particularly in the case of synthetic models where it is not possible to produce outputs that are directly linked to data for example. Very general stylized facts, as aiming at producing an urban hierarchy or a network hierarchy, are relatively limited. Moreover, the hierarchy is mechanically produced by most models including aggregation processes. We therefore need more elaborated indicators to understand the dynamics of the system. These indicators must in particular give elements of answer to the following questions:
\begin{itemize}
	\item types of systems of cities produced by the model;
	\item change in time of the organization of the system of cities;
	\item typical profiles of trajectories;
	\item ability to ``produce some co-evolution''.
\end{itemize}

In order to concentrate on the ability of the model to produce trajectories that are both diverse and complex, and for example its ability to produce bifurcations that would manifest as inversions in ranks, and also its ability to capture different aspects of co-evolutive dynamics, we propose a set of indicators, including for example lagged correlation measures in the spirit of causality regimes introduced by~\cite{raimbault2017identification}, or a correlation measure as a function of distance, to understand the role of spatial interactions in the coupling of trajectories.

\subsubsection{Indicators}

Given a variable $X_i(t)$ defined for each city and in time (that will be the population or centrality measures for example), we define the following indicators.

\paragraph{Summary statistics}

Simple but crucial indicators characterize the distribution of $X_i$ in time:
\begin{itemize}
	\item hierarchy (slope of the least squares adjustment of $X_i$ as a function of rank) $\alpha (t)$
	\item entropy of the distribution $\varepsilon (t)$
	\item descriptive statistics (average $\hat{\mathbb{E}}\left[X\right] (t)$ and standard deviation $\hat{\sigma} (t)$)
\end{itemize}

\paragraph{Rank correlation}

The rank correlation between the initial time and the final time, which translates the quantity of change in the hierarchy during the evolution of the system, and is defined by

\begin{equation}
\rho_r = \hat{\rho}\left[rg(X_i(t=0)),rg(X_i(t=t_f))\right]
\end{equation}

where $rg(X_i)$ is the rank of $X_i$ among all values (similar to a Spearman rank correlation).

\paragraph{Diversity of trajectories}

The diversity of trajectories $\mathcal{D}\left[X_i\right]$ captures a diversity of time series profiles for the considered variable. With $\tilde{X}_i(t)\in \left[0;1\right]$ the trajectories that have been individually rescaled, it is defined by

\begin{equation}
\mathcal{D}\left[X_i\right] = \frac{2}{N\cdot(N-1)}\sum_{i<j} \left(\frac{1}{T}\int_{t} \left(\tilde{X}_i(t) - \tilde{X}_j(t)\right)^2 \right)^{\frac{1}{2}}
\end{equation}

The L2-norm can be generalized by any Minkovski distance, as done in~\cite{raimbault2016hybrid}. More elaborated indices for this aspect could imply the use of specific time-series clustering techniques~\cite{liao2005clustering}.

\paragraph{Shape of trajectories}

We quantify the shape of temporal trajectories through the changes in direction of these $\mathcal{C}\left[X_i\right]$, that we take as the number of local extrema, detected by a change of sign of the derivative. In the context of such a type of model, which mainly produces monotonous trajectories, this indicator witnesses in a certain way of a ``complexity'' of trajectories. In the case of more elaborated shapes, measures such as permutation entropy would be better candidates \cite{scarpino2017predictability}.

\paragraph{Distance correlations}

We also introduce the correlations as a function of distance, to understand the way the effect of distance is translated at the macroscopic scale. The profile of this function, regarding interaction distance parameters included in the model, will translate the tendency of the model to lead to the emergence of one level of interaction or the other. It is computed as

\begin{equation}
\rho_d = \hat{\rho}\left[(X(\vec{x}_k,Y(\vec{x}_{k'}))\right]
\end{equation}

where $X_i, Y_i$ are the two variables considered and $(k,k')$ the set of couples such that $\norm{\vec{x}_k-\vec{x}_{k'}} - d \leq \varepsilon$ with $\varepsilon$ a tolerance threshold (in practice taken to regroup couples by distance deciles).

\paragraph{Lagged correlations}

Lagged correlations between the variations of variables, to identify causality patterns between variables $X$ and $Y$. The patterns $\hat{\rho}_{\tau}$ for all variables, and for $\tau$ lag or anticipation, must be understood in the sense of potential regimes, explored by \cite{raimbault2017identification}.

\begin{equation}
\rho_{\tau} = \hat{\rho}\left[\Delta X(t-\tau),\Delta Y(t)\right]
\end{equation}

\subsubsection{Variables}

These indicators are used in our case on the following variables:
\begin{itemize}
	\item populations $\mu_i(t)$,
	\item closeness centralities
	\[c_i(t) = \frac{1}{N-1}\sum_{i\neq j} \frac{1}{d_{ij}(t)}\]
	which capture the position within the urban system,
	\item accessibilities \[X_i = \frac{1}{\sum_k \mu_k}\sum_{i\neq j} P_j \exp{\left(- d_{ij}(t)/d_G\right)}\] which capture the insertion within the urban system.
\end{itemize}

They capture both city trajectories, network trajectories, and the coupling of both with accessibility. The application of above operators to these state variables will thus inform on trajectories of cities, trajectories of the network and trajectories of their coupling, whereas operators based on correlations will inform on interactions between the two aspects.

\section{Results}

\subsection{Model implementation}

We modified and extended the NetLogo implementation of the model provided by \cite{schmitt2014modelisation}, to include in particular (i) methods for the synthetic setup; (ii) indicators described above; (iii) methods for the inclusion within OpenMOLE experiments. The modified code with exploration scripts are available on the open git repository of the project at \url{https://github.com/JusteRaimbault/CityNetwork/tree/master/Models/Reproduction/SimpopNet}.

\subsection{Experience plan}

Given an initial spatial configuration (i.e. a value of meta-parameters for the initial city system and network generator), we establish the behavior of indicators by exploring a grid of the parameter space. The number of parameters being relatively and the objective being a first grasp of the model behavior, in particular if it is able to produce co-evolution dynamics, we do not use more elaborated exploration methods. The parameters are $(d_G,\gamma_G,\gamma_N,\theta_N,v_0)$ and the meta-parameters $(N_S,\alpha_S,d_S,n_S)$. We take also the meta-parameters into account in order to understand the sensitivity of the model to space.

We explore a grid of 16 configurations of meta-parameters (see Table~\ref{tab:macrocoevolexplo:spacematters} for all values), 324 configurations of parameters (such that $d_G \in \left[0.001, \ldots , 0.016\right]$ by $0.005$, $\gamma_G \in \left[0.5 , \ldots 2.5\right]$ by $1.0$, $\gamma_N \in \left[0.5 , \ldots 2.5\right]$ by $1.0$, $\theta_N \in \left[1.0 , \ldots , 21.0\right]$ by $10.0$ and $v_0 \in \left[10.0, \ldots , 110.0\right]$ by $50.0$), and 30 random replications, what corresponds to $155520$ simulations. They are executed on a computation grid with the intermediary of OpenMole. Simulation results are available at \url{http://dx.doi.org/10.7910/DVN/RW8S36}.

\subsection{Convergence}

Since the model is stochastic, it is important to control the convergence of indicators, that will be more or less easy depending on their variability. To quantify the variability of an indicator $X$ regarding stochasticity, we use a measure similar to the one used by~\cite{raimbault2018calibration}, given by $v\left[X\right] = \hat{\mathbb{E}}\left[X\right]/\hat{\sigma}\left[X\right]$ with basic estimators for the expectancy and the standard deviation. On the full set of replications, we obtain for all indicators given previously, a median for the ratio $v\left[X\right]$ estimated within the 30 replications, estimated on all parameter values, which takes a minimal value of $3.94$, for the average accessibility at final time, what witnesses a low stochastic variability. We can furthermore use this value to estimate the level of convergence: it corresponds to a 95\% confidence interval around the mean of relative size $0.18$ (under the assumption of a normal distribution of the average), i.e. a good convergence. This aspect is crucial for the robustness of results, as this experiment shows that working with this number of repetitions and aggregate averages is consistent.

\subsection{Sensitivity to spatial initial conditions}

We quantify the sensitivity to spatial initial conditions by using the definition of the relative measure of sensitivity, given by~\cite{cottineau2017initial}. This measure is for two phase diagrams $f_1,f_2$ and $d$ euclidian distance, 

\begin{equation}
\tilde{d} = 2 \cdot \frac{d(f_1,f_2)}{(\Varb{f_1}+\Varb{f_2})}
\end{equation}

Table~\ref{tab:macrocoevolexplo:spacematters} gives values of $\tilde{d}$ for the 16 configurations of meta-parameters. These are given in comparison to an arbitrary reference configuration (first column). The hierarchy within the initial system of cities appears as the stronger determinant of variability, since all configurations with $\alpha_S = 1.5$ give values larger than $1.7$, what witnesses a very strong sensitivity relative to this hierarchy.

Then, the number of cities plays a non negligible secondary role, giving the stronger effects of space. Thus, it is crucial to keep in mind this role of the initial configuration during the analysis of phase diagrams. To stay within the same spirit than the model that was initially proposed, we will however comment a phase diagram for a given spatial configuration. The study of the extended model with integration of meta-parameters to which it is sensitive at their full extent is however beyond the reach of this first analysis.

\begin{table}[!ht]
\caption{\textbf{Sensitivity to space of the SimpopNet model.} Each column corresponds to an instance of the phase diagram, for which meta-parameters are given, with the relative distance to an arbitrary reference diagram. As inputs we have the meta-parameters $N_S,\alpha_S,d_S,n_S$ and as outputs of simulations the distance $\tilde{d}$.\label{tab:macrocoevolexplo:spacematters}}
\centering
\begin{tabular}{|l|l|l|l|l|l|l|l|l|l|l|l|l|l|l|l|l|}
\hline
$N_S$ & 40 & 40 & 40 & 40 & 40 & 40 & 40 & 40 & 80 & 80 & 80 & 80 & 80 & 80 & 80 & 80\\
$\alpha_S$ & 0.5 & 0.5 & 0.5 & 0.5 & 1.5 & 1.5 & 1.5 & 1.5 & 0.5 & 0.5 & 0.5 & 0.5 & 1.5 & 1.5 & 1.5 & 1.5\\
$d_S$ & 5 & 5 & 10 & 10 & 5 & 5 & 10 & 10 & 5 & 5 & 10 & 10 & 5 & 5 & 10 & 10\\
$n_S$ & 10 & 30 & 10 & 30 & 10 & 30 & 10 & 30 & 10 & 30 & 10 & 30 & 10 & 30 & 10 & 30\\\hline
$\tilde{d}$ & 0 & 0.05 & 0.26 & 0.21 & 1.79 & 1.80 & 1.79 & 1.72 & 0.44 & 0.36 & 0.42 & 0.42 & 2.25 & 2.23 & 2.24 & 2.21\\\hline
\end{tabular}
\end{table}

\begin{figure}
	\includegraphics[width=\linewidth,height=0.85\textheight]{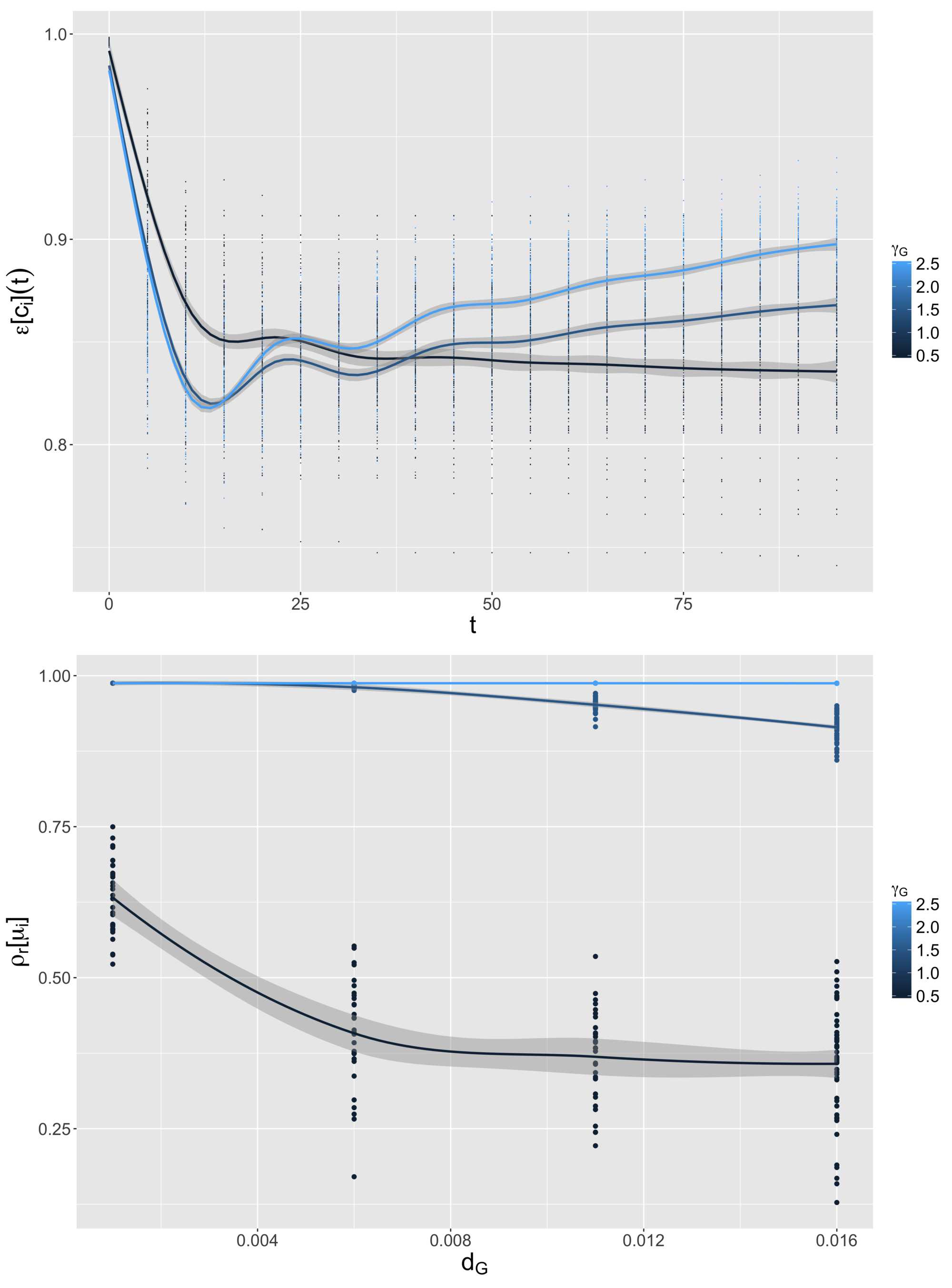}
	\caption{\textbf{Model behavior for the spatial configuration $N_S=80,\alpha_S=0.5,d_S=10,n_S=30$.} (\textit{Top}) Temporal trajectories of the entropy for closeness centralities, for $\gamma_N = 2.5$, $v_0 = 110$, $d_G = 0.016$, $\theta_N = 11$, as a function of $\gamma_G$ (color); (\textit{Bottom}) Rank correlation for population, as a function of $d_G$ and of $\gamma_G$ (color), for $\theta_N = 11$, $\gamma_N = 2.5$.\label{fig:macrocoevolexplo:behavior}}
\end{figure}

\subsection{Model behavior}

The Fig.~\ref{fig:macrocoevolexplo:behavior} reports the behavior of the model according to a selection among the diverse indicators given above. We comment a particular spatial configuration which corresponds to a low hierarchical system with a network having only local shortcuts, given by meta-parameters $N_S=80,\alpha_S=0.5,d_S=10,n_S=30$, which are the values giving configurations that are the most similar to the one of the original model. More exhaustive plots for this parameter configuration are available in~\cite{raimbault:tel-01857741} (Appendix A.7).

The values taken by the entropy for centralities (first panel of Fig.~\ref{fig:macrocoevolexplo:behavior}), as a function of time, for $\gamma_N = 2.5$ and $v_0 = 110$, exhibit different regimes depending on $d_G$ and $\gamma_G$. A low hierarchy leads to an entropy stabilizing in time, what corresponds to a certain uniformization of distances. On the contrary, a strong hierarchy produces a regime with a minimum, and then an increase of disparities in time. More hierarchical interactions produce more hierarchical systems on the long terms, what could have been naturally expected, but with a transient behavior in which the system goes through a point with a maximum of equality between cities in terms of centralities. This confirms that taking into account dynamics in systems of cities is crucial for their understanding.

This variety of behaviors can be found again with the rank correlation $\rho_R$, that we show here for the population variable, as a function of $d_G$. It has a low sensitivity to $\theta_N$ and $\gamma_N$ (see~\cite{raimbault:tel-01857741}, Appendix A.7), but strongly varies as a function of $d_G$ and $\gamma_G$ as shiwn in Fig.~\ref{fig:macrocoevolexplo:behavior} (second panel): interactions at a higher distance induce systematically a larger number of changes in the hierarchy of populations. These can occur when the hierarchy of distance is low. To summarize, the increase of the range of interactions will diminish the inertia of trajectories of the system of cities, whereas the increase of their hierarchy will increase it. This is relatively credible from a thematic point of view: longer and uniform interactions have more chances to make individual trajectories change.

\begin{figure}
\includegraphics[width=\linewidth,height=0.9\textheight]{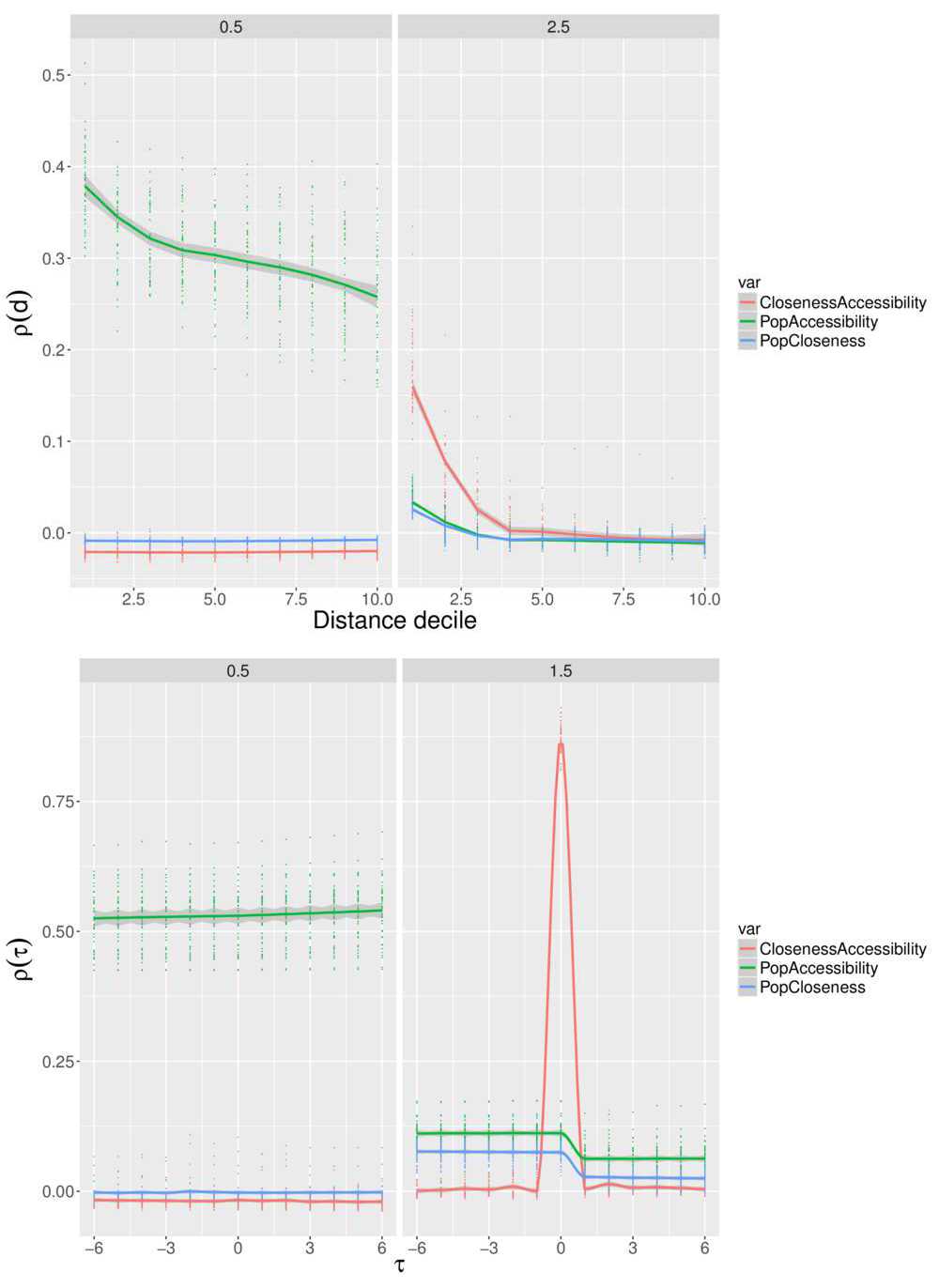}
	\caption{\textbf{Correlations in the model for the spatial configuration $N_S=80,\alpha_S=0.5,d_S=10,n_S=30$.} (\textit{Top}) Correlations as a function of distance, for couples of variables (color), for $\gamma_N = 2.5$, $\theta_N = 21$, $v_0 = 10$, and for $d_G$ (columns) and $\gamma_G$ (rows) variables; (\textit{Bottom}) Lagged correlations for the same parameters. \label{fig:macrocoevolexplo:correlations}}
\end{figure}

The behavior of correlation indicators is shown in Fig.~\ref{fig:macrocoevolexplo:correlations}. Concerning the effect of distance on correlations between variables, i.e. the evolution of $\rho_d$, it is interesting to note that an increase of $d_G$ systematically diminishes the levels of correlation, what corresponds to the complexification that we previously showed. As expected, $\rho_d\left[d\right]$ decreases as a function of distance, and exhibits non zero values for the correlation between population and centrality for a high hierarchy $\gamma_G$, what shows that simultaneous adaptation regimes are rare in this model.

\subsection{Causality regimes}

Finally, by studying $\rho_{\tau}$ (Fig.~\ref{fig:macrocoevolexplo:correlations}, bottom panel), we observe that causality regimes in the sense of~\cite{raimbault2017identification} are relatively restrained (see~\cite{raimbault:tel-01857741}, Appendix A.7, for the confirmation for a broader range of parameters). The population is systematically caused by the centrality, but there exists no regime in which we observe the contrary. This is a logic of an effect of reinforcement of hierarchy by centrality. This exploration does not provide a configuration with circular causalities, and thus not a co-evolution properly speaking as we defined in the statistical sense.




\subsection{PSE algorithm}

\begin{figure}
	\includegraphics[width=\textwidth]{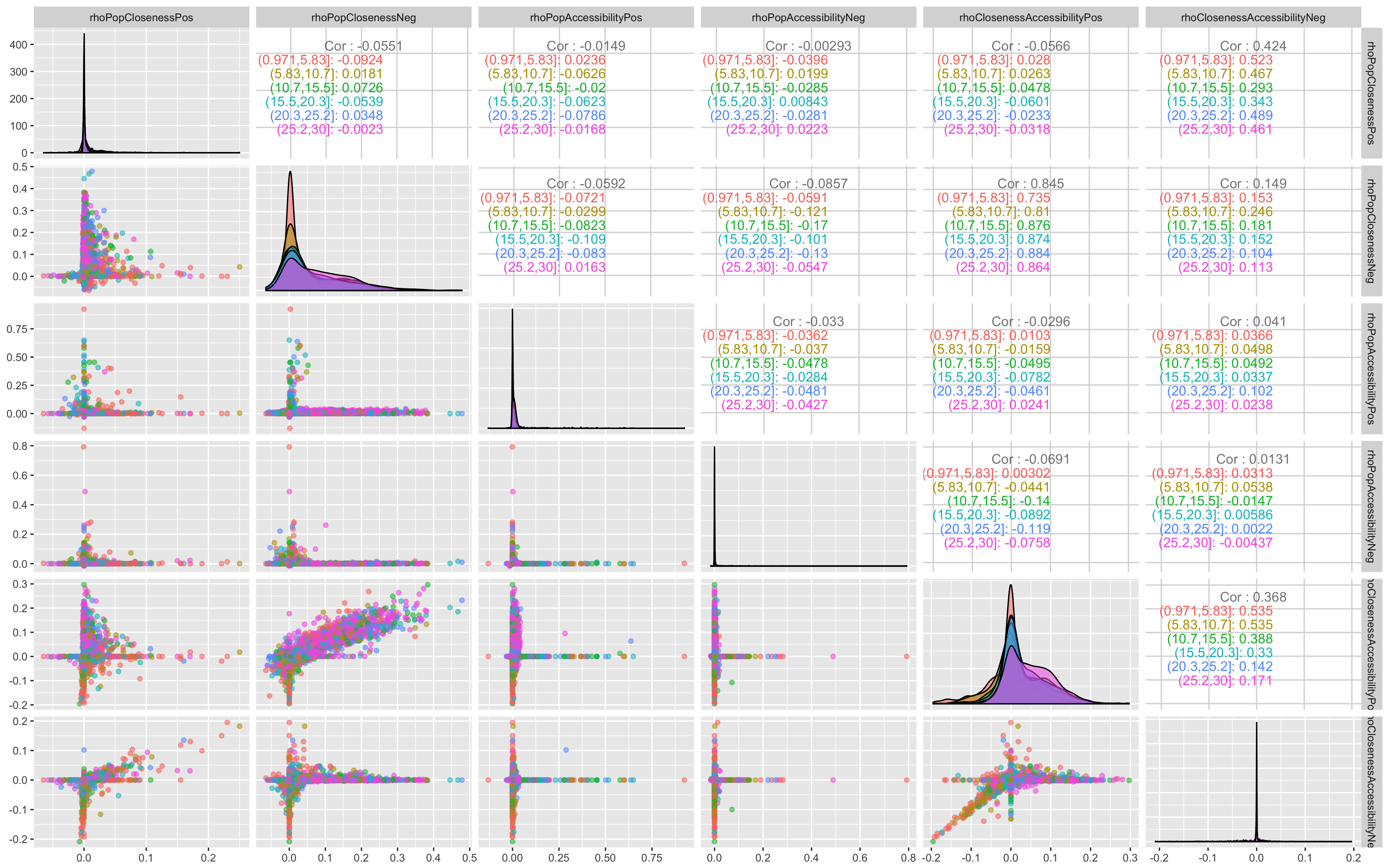}
	\caption{\textbf{Feasible space of lagged correlation obtained with the PSE algorithm.} We give as a scatterplot the six objectives of the algorithm. Color level gives $\theta_N$, which does not seems to be a simple driver of correlation values.\label{fig:pse}}
\end{figure}

The last conclusion is crucial regarding the thematic questions that one can ask to the model, and was obtained with a limited experiment (simple grid sampling). However, because precisely of non-linearities, such simple sampling may miss regions of the parameter space in which dramatic changes occurs in the phase diagram of the model, as \cite{cherel2015beyond} showed on toy models and on the Marius model. We apply here therefore the Pattern Space Exploration algorithm, which is precisely designed to unveil such unexpected behavior, what can be abstracted as a sampling of the image space of the model instead of its parameter space.

We use here the following indicators as targets on which the algorithm must find diverse patterns. Given all couples of variables $(X,Y)$ and for $\tau > 0$ and $\tau < 0$, we consider $\max_{\tau} \rho_{\tau}\left[X,Y\right]$ and $\min_{\tau} \rho_{\tau}\left[X,Y\right]$, and report the one with the largest absolute value which absolute value is larger than correlation at the origin, and 0 otherwise. This adaptation of the method of \cite{raimbault2017identification} has been proposed by \cite{2018arXiv180409430R} for a similar models, for which simultaneous correlations are high as a consequence of model structure. The algorithm was run on 100 islands, with 500 parallel instances, and stopped at generation 16000.

The Fig.~\ref{fig:pse} gives the results as a scatterplot of diversity targets of the algorithm. Several regimes are located on the axis, corresponding to one-directional relations between the variables. We obtain however a point cloud within positive correlations for both negative and positive delays for population and centrality, what actually witnesses a co-evolution (circular causality between the two). We have a similar pattern in negative correlations for centrality and accessibility, the point cloud being however more constrained on the diagonal, what may due to a direct repercussion of one variable on the other as they are structurally linked. We also observe points with a positive correlation between population and centrality for positive delays and negative delays, what corresponds to a direct circular causality, and between population and centrality and centrality and accessibility. This analysis reveals thus that the model is actually able to produce co-evolutive regimes in the sense of \cite{raimbault:tel-01857741}.

Although the algorithm seems to have converged (less patterns discovered after generation 10000), a more thorough investigation of the role of stochasticity would be needed since several points have few repetitions, to ensure the robustness of these results. These remain out of the scope of this proof-of-concept exploration, in which we show how a qualitative change can occur in knowledge about a model when using specific exploration methods.

\section{Discussion}


First of all, some thematic observations about the question of interactions between networks and territories can be formulated.


This model could be a useful tool to study the ``tunnel effect'', which is as we recall is the absence of interaction of an infrastructure traversing a territory with it \cite{raimbault2018indirect}. Indeed, the rules allowing variable values for $v(t)$ and the non-planarity mechanism (when a new link is constructed, it does create intersections only with links of similar speed), allows the introduction of this effect. This remains however exogenous since explicitly specified in model rules, on the contrary to the interaction model of \cite{raimbault2018indirect} with feedback of flows, in which the variations of parameters capture an endogenous tunnel effect. The introduction of specific indicators to measure it would be an interesting development direction in the case of this model.



This model could be calibrated on real systems, but the use of the interaction model without the endogenous Gibrat term would be difficultly adaptable to an application of the model on real data because of the values for calibrated for endogenous growth for example by \cite{raimbault2018indirect}. The application to a real system would thus require to study the complete model first.



In comparison, the co-evolution model introduced by \cite{2018arXiv180409430R} seems to be less constrained for network dynamics and to produce more varied interaction regimes. We can not however compare the two in such different contexts. Future work to better understand the role of co-evolution in urban systems shall require multi-modeling approaches \cite{cottineau2015modular} and more systematic model benchmarks.


Regarding the modeling side, this work provides a supplementary proof-of-concept of the importance of the use of new tools and methods to extract knowledge from simulation models, since we indeed showed that the conclusion on the ability of the SimpopNet model to produce co-evolutive regimes would not have been obtained without the use of the PSE algorithm.

\section{Conclusion}


This chapter has illustrated the systematic exploration of a simulation model for the co-evolution of cities and transportation networks. Our contribution covers several dimensions, including a proof-of-concept of the importance of new model exploration tools and methods, a set of indicators to study similar models for systems of cities, and thematic knowledge about the model such as its sensitivity to spatial initial conditions and its ability to produce co-evolutive regimes.

\begin{acknowledgement}
Results obtained in this paper were computed on the vo.complex-system.eu virtual organization of the European Grid Infrastructure ( http://www.egi.eu ). We thank the European Grid Infrastructure and its supporting National Grid Initiatives (France-Grilles in particular) for providing the technical support and infrastructure. This work is part of DynamiCity, a FUI project funded by BPI France, Auvergne-Rhône-Alpes region, Ile-de-France region and Lyon metropolis. This work was also funded by the Urban Dynamics Lab grant EPSRC EP/M023583/1.
\end{acknowledgement}


\begin{thebibliography}{}

\bibitem[Achibet et~al., 2014]{achibet2014model}
Achibet, M., Balev, S., Dutot, A., and Olivier, D. (2014).
\newblock A model of road network and buildings extension co-evolution.
\newblock {\em Procedia Computer Science}, 32:828--833.

\bibitem[Arduin, 2018]{arduin2018modelisation}
Arduin, H. (2018).
\newblock {\em Mod{\'e}lisation math{\'e}matique des interactions entre
  pathog{\`e}nes chez l’h{\^o}te humain: Application aux virus de la grippe
  et au pneumocoque}.
\newblock PhD thesis, Universit{\'e} Paris-Saclay.

\bibitem[Arthur, 2015]{arthur2015complexity}
Arthur, W.~B. (2015).
\newblock Complexity and the shift in modern science.
\newblock Conference on Complex Systems, Tempe, Arizona.

\bibitem[Banos, 2013]{banos2013pour}
Banos, A. (2013).
\newblock {\em Pour des pratiques de mod{\'e}lisation et de simulation
  lib{\'e}r{\'e}es en G{\'e}ographie et SHS}.
\newblock PhD thesis, Universit{\'e} Paris 1 Panth{\'e}on Sorbonne.

\bibitem[Banos, 2017]{banos2017knowledge}
Banos, A. (2017).
\newblock Knowledge accelerator' in geography and social sciences: Further and
  faster, but also deeper and wider.
\newblock In {\em Urban Dynamics and Simulation Models}, pages 119--123.
  Springer.

\bibitem[Baptiste, 2010]{baptistemodeling}
Baptiste, H. (2010).
\newblock Modeling the evolution of a transport system and its impacts on a
  french urban system.
\newblock {\em Graphs and Networks: Multilevel Modeling, Second Edition}, pages
  67--89.

\bibitem[Batty, 2007]{batty2007cities}
Batty, M. (2007).
\newblock {\em Cities and complexity: understanding cities with cellular
  automata, agent-based models, and fractals}.
\newblock The MIT press.

\bibitem[Benenson and Torrens, 2004]{benenson2004geosimulation}
Benenson, I. and Torrens, P.~M. (2004).
\newblock Geosimulation: object-based modeling of urban phenomena.
\newblock {\em Computers, Environment and Urban Systems}, 28(1-2):1--8.

\bibitem[Bird, 2011]{bird2011computing}
Bird, I. (2011).
\newblock Computing for the large hadron collider.
\newblock {\em Annual Review of Nuclear and Particle Science}, 61:99--118.

\bibitem[Blumenfeld-Lieberthal and Portugali, 2010]{blumenfeld2010network}
Blumenfeld-Lieberthal, E. and Portugali, J. (2010).
\newblock Network cities: A complexity-network approach to urban dynamics and
  development.
\newblock In {\em Geospatial Analysis and Modelling of Urban Structure and
  Dynamics}, pages 77--90. Springer.

\bibitem[Brasebin et~al., 2017]{brasebin2017apports}
Brasebin, M., Chapron, P., Ch{\'e}rel, G., Leclaire, M., Lokhat, I., Perret,
  J., and Reuillon, R. (2017).
\newblock Apports des m{\'e}thodes d'exploration et de distribution
  appliqu{\'e}es {\`a} la simulation des droits {\`a} b{\^a}tir.
\newblock In {\em Spatial Analysis and GEOmatics 2017}.

\bibitem[Bretagnolle, 2003]{bretagnolle2003vitesse}
Bretagnolle, A. (2003).
\newblock Vitesse et processus de s{\'e}lection hi{\'e}rarchique dans le
  syst{\`e}me des villes fran{\c{c}}aises.
\newblock {\em Donn{\'e}es urbaines}, 4.

\bibitem[Bretagnolle, 2009]{bretagnolle:tel-00459720}
Bretagnolle, A. (2009).
\newblock {\em {Villes et r{\'e}seaux de transport : des interactions dans la
  longue dur{\'e}e, France, Europe, {\'E}tats-Unis}}.
\newblock Hdr, Universit{\'e} Panth{\'e}on-Sorbonne - Paris I.

\bibitem[Ch{\'e}rel et~al., 2015]{cherel2015beyond}
Ch{\'e}rel, G., Cottineau, C., and Reuillon, R. (2015).
\newblock Beyond corroboration: Strengthening model validation by looking for
  unexpected patterns.
\newblock {\em PloS one}, 10(9):e0138212.

\bibitem[Cottineau et~al., 2017]{cottineau2017initial}
Cottineau, C., Raimbault, J., Le~Texier, M., Le~N{\'e}chet, F., and Reuillon,
  R. (2017).
\newblock Initial spatial conditions in simulation models: the missing leg of
  sensitivity analyses?
\newblock In {\em 2017 International Conference on GeoComputation: Celebrating
  21 Years of GeoComputation}.

\bibitem[Cottineau et~al., 2015]{cottineau2015modular}
Cottineau, C., Reuillon, R., Chapron, P., Rey-Coyrehourcq, S., and Pumain, D.
  (2015).
\newblock A modular modelling framework for hypotheses testing in the
  simulation of urbanisation.
\newblock {\em Systems}, 3(4):348--377.

\bibitem[Epstein, 2006]{epstein2006generative}
Epstein, J.~M. (2006).
\newblock {\em Generative social science: Studies in agent-based computational
  modeling}.
\newblock Princeton University Press.

\bibitem[Galam, 2008]{galam2008sociophysics}
Galam, S. (2008).
\newblock Sociophysics: A review of galam models.
\newblock {\em International Journal of Modern Physics C}, 19(03):409--440.

\bibitem[LeCun et~al., 2015]{lecun2015deep}
LeCun, Y., Bengio, Y., and Hinton, G. (2015).
\newblock Deep learning.
\newblock {\em nature}, 521(7553):436.

\bibitem[Liao, 2005]{liao2005clustering}
Liao, T.~W. (2005).
\newblock Clustering of time series data—a survey.
\newblock {\em Pattern recognition}, 38(11):1857--1874.

\bibitem[Mantegna and Stanley, 1999]{mantegna1999introduction}
Mantegna, R.~N. and Stanley, H.~E. (1999).
\newblock {\em Introduction to econophysics: correlations and complexity in
  finance}.
\newblock Cambridge university press.

\bibitem[Mimeur et~al., 2017]{mimeur:hal-01616746}
Mimeur, C., Queyroi, F., Banos, A., and Th{\'e}venin, T. (2017).
\newblock {Revisiting the structuring effect of transportation infrastructure:
  an empirical approach with the French Railway Network from 1860 to 1910}.
\newblock {\em {Historical Methods: A Journal of Quantitative and
  Interdisciplinary History}}.

\bibitem[Passerat-Palmbach et~al., 2017]{passerat2017reproducible}
Passerat-Palmbach, J., Reuillon, R., Leclaire, M., Makropoulos, A., Robinson,
  E.~C., Parisot, S., and Rueckert, D. (2017).
\newblock Reproducible large-scale neuroimaging studies with the openmole
  workflow management system.
\newblock {\em Frontiers in neuroinformatics}, 11:21.

\bibitem[Paulus, 2004]{paulus2004coevolution}
Paulus, F. (2004).
\newblock {\em Co{\'e}volution dans les syst{\`e}mes de villes: croissance et
  sp{\'e}cialisation des aires urbaines fran{\c{c}}aises de 1950 {\`a} 2000}.
\newblock PhD thesis, Universit{\'e} Panth{\'e}on-Sorbonne-Paris I.

\bibitem[Pumain, 1997]{pumain1997pour}
Pumain, D. (1997).
\newblock Pour une th{\'e}orie {\'e}volutive des villes.
\newblock {\em L'Espace g{\'e}ographique}, pages 119--134.

\bibitem[Pumain, 2010]{pumain2010theorie}
Pumain, D. (2010).
\newblock Une th{\'e}orie g{\'e}ographique des villes.
\newblock {\em Bulletin de la Soci{\'e}t{\'e} g{\'e}ographie de Li{\`e}ge},
  55:5--15.

\bibitem[Pumain, 2014]{pumain2014effets}
Pumain, D. (2014).
\newblock Les effets structurants ou les raccourcis de l'explication
  g{\'e}ographique.
\newblock {\em Espace g{\'e}ographique}, 43(1):65--67.

\bibitem[Pumain and Reuillon, 2017a]{pumain2017evaluation}
Pumain, D. and Reuillon, R. (2017a).
\newblock Evaluation of the simpoplocal model.
\newblock In {\em Urban Dynamics and Simulation Models}, pages 37--56.
  Springer.

\bibitem[Pumain and Reuillon, 2017b]{pumain2017urban}
Pumain, D. and Reuillon, R. (2017b).
\newblock {\em Urban Dynamics and Simulation Models}.
\newblock Springer International.

\bibitem[Raimbault, 2016a]{raimbault2016cautious}
Raimbault, J. (2016a).
\newblock For a cautious use of big data and computation.
\newblock In {\em Royal Geographical Society-Annual Conference 2016-Session:
  Geocomputation, the Next 20 Years (1)}.

\bibitem[Raimbault, 2016b]{raimbault2016generation}
Raimbault, J. (2016b).
\newblock G{\'e}n{\'e}ration de donn{\'e}es synth{\'e}tiques corr{\'e}l{\'e}es.
\newblock In {\em Rochebrune 2016, Journ{\'e}es d'Etude sur les Syst{\`e}mes
  Complexes Naturels et Artificiels}.

\bibitem[Raimbault, 2017a]{raimbault2017applied}
Raimbault, J. (2017a).
\newblock An applied knowledge framework to study complex systems.
\newblock In {\em Complex Systems Design \& Management}, pages 31--45.

\bibitem[Raimbault, 2017b]{raimbault2017identification}
Raimbault, J. (2017b).
\newblock Identification de causalités dans des données spatio-temporelles.
\newblock In {\em Spatial Analysis and GEOmatics 2017}.

\bibitem[Raimbault, 2017c]{raimbault2017models}
Raimbault, J. (2017c).
\newblock Models coupling urban growth and transportation network growth: An
  algorithmic systematic review approach.
\newblock {\em Plurimondi}, (17).

\bibitem[{Raimbault}, 2018]{2018arXiv180505195R}
{Raimbault}, J. (2018).
\newblock {An Urban Morphogenesis Model Capturing Interactions between Networks
  and Territories}.
\newblock {\em ArXiv e-prints}.

\bibitem[Raimbault, 2018a]{raimbault2018calibration}
Raimbault, J. (2018a).
\newblock Calibration of a density-based model of urban morphogenesis.
\newblock {\em PLoS ONE, in press}.

\bibitem[Raimbault, 2018b]{raimbault:tel-01857741}
Raimbault, J. (2018b).
\newblock {\em {Characterizing and modeling the co-evolution of transportation
  networks and territories}}.
\newblock Theses, {Univerist{\'e} Paris 7 Denis Diderot}.

\bibitem[Raimbault, 2018c]{raimbault2018indirect}
Raimbault, J. (2018c).
\newblock Indirect evidence of network effects in a system of cities.
\newblock {\em Environment and Planning B: Urban Analytics and City Science},
  page 2399808318774335.

\bibitem[{Raimbault}, 2018]{2018arXiv180409430R}
{Raimbault}, J. (2018).
\newblock {Modeling the co-evolution of cities and networks}.
\newblock {\em ArXiv e-prints}.

\bibitem[Raimbault et~al., 2014]{raimbault2016hybrid}
Raimbault, J., Banos, A., and Doursat, R. (2014).
\newblock A hybrid network/grid model of urban morphogenesis and optimization.
\newblock In {\em 4th International Conference on Complex Systems and
  Applications}, pages 51--60.

\bibitem[Reuillon et~al., 2013]{reuillon2013openmole}
Reuillon, R., Leclaire, M., and Rey-Coyrehourcq, S. (2013).
\newblock Openmole, a workflow engine specifically tailored for the distributed
  exploration of simulation models.
\newblock {\em Future Generation Computer Systems}, 29(8):1981--1990.

\bibitem[Reuillon et~al., 2015]{reuillon2015new}
Reuillon, R., Schmitt, C., De~Aldama, R., and Mouret, J.-B. (2015).
\newblock A new method to evaluate simulation models: the calibration profile
  (cp) algorithm.
\newblock {\em Journal of Artificial Societies and Social Simulation},
  18(1):12.

\bibitem[Scarpino and Petri, 2017]{scarpino2017predictability}
Scarpino, S.~V. and Petri, G. (2017).
\newblock On the predictability of infectious disease outbreaks.
\newblock {\em arXiv preprint arXiv:1703.07317}.

\bibitem[Schamp, 2010]{schamp201020}
Schamp, E.~W. (2010).
\newblock 20 on the notion of co-evolution in economic geography.
\newblock {\em The handbook of evolutionary economic geography}, page 432.

\bibitem[Schmitt, 2014]{schmitt2014modelisation}
Schmitt, C. (2014).
\newblock {\em Mod{\'e}lisation de la dynamique des syst{\`e}mes de peuplement:
  de SimpopLocal {\`a} SimpopNet.}
\newblock PhD thesis, Universit{\'e} Panth{\'e}on-Sorbonne-Paris I.

\bibitem[Wal and Boschma, 2011]{doi:10.1080/00343400802662658}
Wal, A. L. J.~T. and Boschma, R. (2011).
\newblock Co-evolution of firms, industries and networks in space.
\newblock {\em Regional Studies}, 45(7):919--933.

\end{thebibliography}
\end{document}